\documentclass{optica-article}

\journal{opticajournal} 

\articletype{Research Article}
\usepackage{graphicx,caption}
\usepackage{lineno}
\usepackage[upgreek]{mathastext}
\usepackage{textgreek}
\usepackage{upgreek}
\begin{document}

\title{DC-to-GHz Modulation In Microring Modulators Using Ferroelectric Nematic Liquid Crystal-on-Silicon in a Foundry Photonic Process}

\author{Iman Taghavi,\authormark{1,2,*} Alexander Tofini,\authormark{1} Edward K. Leung,\authormark{1} Cory Pecinovsky,\authormark{3} Nicolas A.F. Jaeger,\authormark{1} Lukas Chrostowski,\authormark{1,2,4} and Sudip Shekhar \authormark{1,2}}

\address{\authormark{1}Department of Electrical and Computer Engineering, University of British Columbia, 2332 Main Mall, Vancouver, BC V6T 1Z4, Canada\\
\authormark{2}Dream Photonics Inc., 2366 Main Mall, BC V6T 1Z4, Vancouver, Canada\\
\authormark{3}Polaris Electro-Optics, 3400 Industrial Ln, 3130 25th St., Broomfield, CO 80020 USA\\
\authormark{4}Quantum Matter Institute, University of British Columbia, Vancouver, 2355 E Mall, Vancouver, BC V6T 1Z4, Canada\\}   

\email{\authormark{*}staghavi3@ece.ubc.ca; https://orcid.org/0000-0002-2955-9209} 


\begin{abstract*}
	 Silicon-organic hybrid (SOH) platforms exhibit exceptional electro-optic (EO) properties, including high-speed operation, low energy consumption, and compact footprints. However, the absence of a scalable poling method for EO polymers, combined with the slow switching speeds characteristic of liquid crystals, has impeded the integration and compatibility of these materials with commercial silicon photonic foundries. On the other hand, the realization of very-large-scale photonic integrated circuits (PICs) in the native silicon photonics platform itself is impeded by the complexities associated with the wavelength and thermal stabilization for microring modulators (MRMs). This study establishes the foundation for a poling-free, CMOS-compatible SOH MRM platform by exploiting simultaneous AC phase shifts in ferroelectric nematic liquid crystals (FN-LCs). We present the first demonstration of an MRM coated with FN-LC, with both the RF signal and the DC bias applied to the same electrodes, taking advantage of the dual phase shift uniquely available in FN-LC. An EO bandwidth of \textit{f}$_{-6dB}\approx7.8~GHz$ is achieved \textendash~ the highest reported value for an SOH MRM to date. As a proof of concept, we demonstrate an approximately linear resonance shift across the full width at half maximum ($\approx$ 150 pm/V), with a static power efficiency of $\approx$ 4.5 nW/$\pi$ for an MRM occupying a total footprint of $\approx 0.084~\textit{mm}^2$ and exhibiting an on-chip optical insertion loss of $\approx$ 0.78 dB. Successful infiltration of FN-LC, selectively patterned on top of phase shifters, along with optical input/output channels established using free-form photonic wire bonds, is demonstrated in the proposed PIC.
\end{abstract*}

\section{Introduction}

The number of optical components in photonic integrated circuits (PICs) has been steadily growing towards a very-large-scale integration (VLSI)~\cite{shekhar2024roadmapping}. Microring resonators (MRRs) and modulators (MRMs) offer a small footprint, making them ideally suited for dense integration. Still, the need for thermal tuners for control and stabilization adds complexity to routing DC and radio-frequency (RF) interconnects and introduces thermal crosstalk. Thus, the component density is severely affected, and power consumption increases, limiting VLSI scaling and the realization of complex photonic systems for communications, computing, and switching.

Pockels-based silicon photonics can outperform carrier-based phase shift mechanisms (e.g., thermo-optic \cite{yang2018general} and plasma dispersion/free-carrier effects \cite{han2023slow}) PIC building blocks such as modulators \cite{tahersima2019coupling}, switches \cite{enami2011short}, and weight banks \cite{nezami2022packaging}.
Although the effect has been demonstrated in inorganic materials such as lithium niobate \cite{wang2018integrated} and barium titanate \cite{abel2013strong}, parallel advancements in organic electro-optic (EO) materials \textemdash~primarily EO polymers (EOP) \cite{kieninger2018ultra}
\textemdash~have achieved several device performance metrics that surpass those of their inorganic counterparts \cite{taghavi2022polymer, burla2019500, koeber2015femtojoule}.\par


EOP-based devices have seen significant advancements, particularly in material development, enabling their potential integration within large-scale CMOS foundries. Decomposition temperatures as high as $317^{\circ}$C \cite{xu2020ultrahigh} and stable performance at operating temperatures up to $110^{\circ}$C \cite{lu2020high} have been reported. Despite these improvements, the intrinsic thermal sensitivity of poled chromophores above their glass transition temperature ($\textit{T}_g$) remains a challenge that must be addressed \cite{hammond2022organic}. As most reported $\textit{T}_g$ values fall within the $85\sim175^{\circ}$C range \cite{xu20214}, the poling process cannot be integrated into the majority of nanofabrication steps, except during the back-end-of-line post-processing. Large-scale poling is further complicated by leakage current through (LCT) \cite{huang2010highly,jin2014benzocyclobutene}, typically necessitating the inclusion of a charge barrier layer (CBL). More critically, the selective poling or re-poling of EOP-covered devices has yet to be demonstrated, due to the indiscriminate effect of high-temperature electro-thermal poling on previously poled regions. Alternative poling techniques, such as corona discharge \cite{hill1994corona} and laser-assisted methods \cite{donval1999comparative}, often require complex instrumentation and are less efficient. To mitigate premature degradation, poling must be carried out under strictly controlled environmental conditions, including precise regulation of temperature, oxygen, and humidity. Furthermore, several additional fabrication steps are necessary to ensure proper device performance. These include overetching of the underlying oxide layer, surface passivation using metal oxides, and vacuum-assisted curing to fully infiltrate the EOP and achieve a homogeneous, void-free coating \cite{taghavi2022enhanced,gould2011silicon}. Collectively, these technical hurdles have limited the broader adoption of EOP-based PICs by mainstream silicon photonics foundries.\par   

Paraelectric nematic liquid crystal (PN-LC)-based devices have also demonstrated promising performance, owing to their high susceptibilities arising from phenomena such as birefringence \cite{xing2015digitally} and the thermo-optic effect \cite{wang2021thermo}. Phase shifts in PN-LC devices are achieved by rotating the molecules at room temperature, resulting in a significant change in the index and enabling very short phase shifters and low voltages. This poling-free characteristic makes them well-suited for large-scale integration, where numerous PN-LC-coated phase shifters must operate with minimal crosstalk. However, their limited responsiveness to rapidly varying electrical stimuli constrains the EO bandwidth of such devices to the tens of megahertz range \cite{geis201030,wang2014electrically,notaros2022integrated}.\par

A novel state of matter, the ferroelectric nematic liquid crystal (FN-LC), has recently been introduced \cite{lavrentovich2020ferroelectric,chen2020first,kumari2023ferroelectric}. Second harmonic generation measurements and field-induced polarization reversal current have confirmed a non-zero second-order susceptibility ($\upchi^{(2)}$) and a tremendous spontaneous polarization ($\textit{P}_s$), respectively \cite{chen2020first}. The amplitude of the resultant EO effect due to this $\upchi^{(2)} \neq 0$ is attributed to its highly polarizable molecular slow axis, which aligns with both the director and the polar axis \cite{folcia2022ferroelectric}. Similar to ferroelectric crystals, but unlike EOPs, the polar alignment is spontaneous. Since any external field will couple with the macroscopic polarization of the FNLC, very low external electric fields ($\textit{E}_{ext}$) can be used to set the orientation of the polar axis. This property is particularly advantageous for advanced modulation schemes and coherent data transmission, where avoiding a power-intensive thermoelectric heater is beneficial. In specific applications that leverage ring resonators, a full free spectral range (FSR) detuning is required to tune the ring resonance to the operating wavelength.\par

Switching speeds on the order of $\mu$s have been demonstrated with FN-LCs, which is several orders faster than PN-LC \cite{chen2020first}. In fact, the presence of both EO mechanisms (i.e., Pockels and molecular orientation birefringence) makes them truly unique. Recent demonstrations involving doped finger subwavelength gratings \cite{taghavi2025ghz} and conventional slot waveguides \cite{onural2024hybrid,chiang2025ferroelectric-arxiv,chiang2025ferroelectric-IPC} coated with FN-LC substantiate the material's potential. Nevertheless, both implementations have employed a Mach-Zehnder modulator (MZM) configuration showing moderate modulation efficiencies, primarily limited by the low EO coefficient ($\textit{r}_\text{33}$) of FN-LC at RF frequencies compared to an EOP-based counterpart.\par  

This work introduces the first implementation of an MRM based on FN-LC. The material provides two distinct phase-shifting mechanisms across different operating frequencies and exhibits unique EO properties, including inherent self-protection against aging and high viscosity suitable for localized deposition. To address the aforementioned limitations of existing SOH technologies, namely, the poling dependency of EOP-based devices and the limited speed of PN-LC-based devices, the fabricated MRM integrates these material advantages with improved detuning and modulation efficiencies. The intrinsic DC detuning capability and poling-free operation of FN-LC are emphasized as practical solutions to long-standing challenges in SOH MRMs. To demonstrate the scalability and CMOS compatibility of the proposed FN-LC platform, a proof-of-concept PIC was fabricated through a commercial foundry and packaged using both electrical and photonic wire bonding (PWB). 

\section{Design}
The choice of the active material integrated with the photonic backbone (e.g., Si, SiN, or InP) is pivotal in determining the fabrication complexity of the resulting hybrid platform. For organic EO (OEO) platforms, the associated fabrication steps are typically straightforward. The liquid-phase nature of OEO materials enables direct deposition, either manually or via scalable techniques such as inkjet printing \cite{zhang2023recent}, thereby allowing efficient infiltration into nanostructures and enhancing interaction with confined optical modes.\par

\begin{figure}[ht!]
\centering\includegraphics[width=0.8\textwidth]{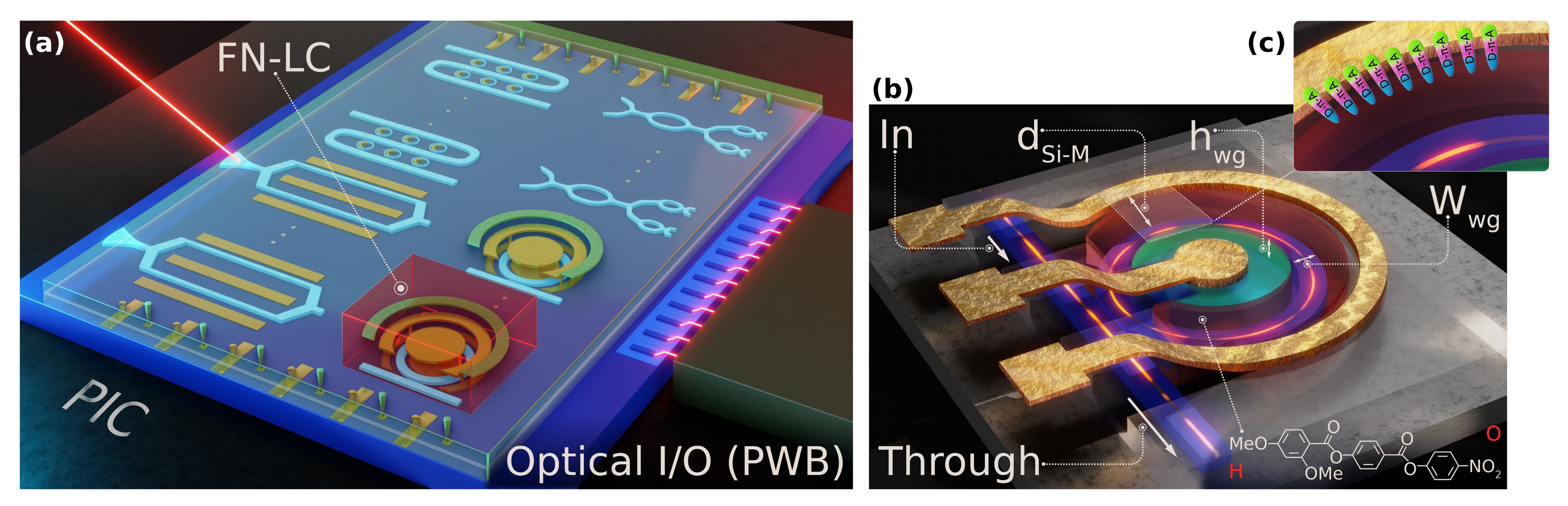}
\caption{Concept of a photonic integrated circuit (PIC) based on a ferroelectric nematic liquid crystal hybrid (FN-LC). (a) The PIC includes stand-alone and nested ring modulators (MRM) and Mach-Zehnder interferometers that can operate as modulators, weight banks, filters, interconnects, and switches. The material is applied selectively to active components, leaving other surface features unaffected. Subwavelength grating couplers connect individual devices to establish device-level optical I/O. An \textit{n}-channel fiber array co-packaged with the PIC connects light via photonic wire bonds (PWBs) produced using two-photon lithography. (b) An MRM example with a clad opening for FN-LC infiltration into the light-matter interaction area space of the waveguide. The metal-to-silicon distance ($\textit{d}_{Si-M}$), as well as the waveguide's width ($\textit{W}_\text{wg}$) and height ($\textit{h}_\text{wg}$), can all be tuned to make a trade-off between drive voltage and electro-optic bandwidth. 
(c) A zoomed-in view of the material's donor-bridge-acceptor structure ("D-$\uppi$-A") rotated in response to an external field, causing the molecular director to get aligned along the spontaneously produced polarization.}
\label{overview}
\end{figure}

Distinct material characteristics of FN-LC—particularly high viscosity at least at room temperature—can be found advantageous to selectively coat active components within a representative PIC comprising wavelength-selective banks, MRMs, and MZMs, as illustrated in Fig.~\ref{overview}. Additionally, FN-LC materials do not require extensive fume-extraction infrastructure to manage outgassing during processing. Specifically, Fig.~\ref{overview}(b) demonstrates how a standard silicon ring resonator can be functionally activated by infiltrating FN-LC into the nanogaps where strong light-matter interaction (LMI) occurs.\par  

The efficiency of a given SOH-based device can be maximized by adopting improved LMI offered in waveguide architectures, such as slot waveguides, where the optical mode, RF electric field, and OEO are all confined in the same physical domain \cite{almeida2004guiding}. A figure of merit for the efficiency of a Pockels-based EO phase shifter ($\eta_{ps}$) given by \cite{taghavi2022polymer}  

\begin{equation}
	\eta_{ps}=\frac{\partial n}{\partial V}=\frac{1}{2}\big[\textit{n}_{eo}^3\textit{r}_{33}\big]\times\big[\frac{\Gamma}{d}\big]
\end{equation}

\noindent where $\textit{n}$ and $\textit{V}$ denote the effective index of the waveguide and the applied voltage, respectively. The first bracket represents the material factor, encompassing the in-device refractive index ($\textit{n}_{eo}$) and the in-device EO coefficient ($\textit{r}_{33}$). The second bracket corresponds to the device factor, which includes the electrode separation ($\textit{d}$) and the field overlap integral ($\Gamma$) \cite{witzens2010design}. Due to the discontinuity of the electric field, slot waveguides exhibit stronger optical mode confinement, enabling a high $\Gamma$ over the reduced distance $\textit{d}$, achieved through appropriate doping profiles \cite{zwickel2020verified}. While slot waveguides exhibit superior $\Gamma/d$ ratios, their effective $\textit{r}_{33}$-and thus the overall material factor-is typically lower than that of thin-film configurations. This reduction arises from LCT during the poling process and the limited attainable acentric molecular order ($<\cos^3\theta>$) in nanostructured environments \cite{heni2017nonlinearities}. Surface passivation using atomic layer deposition of TiO$_2$ or Al$_2$O$_3$ has been shown to serve as a CBL, enhance the uniformity of the poling field at the electrode/EO material interface, and improve $<\cos^3\theta>$ \cite{taghavi2022enhanced,schulz2015mechanism}.\par

While increasing the slot width can partially mitigate these limitations and improve compatibility with CMOS fabrication constraints, it contradicts the design principle of slot waveguides. Alternative structures with enhanced LMI, such as one-dimensional \cite{inoue2013electro} and two-dimensional photonic crystal \cite{zhang2013wide} waveguides, thinned configurations \cite{lu2020high}, and plasmonic slot designs \cite{heni2017silicon}-offer high $\eta_{ps}$ but often lack compatibility with foundry CMOS SOI processes. In contrast, non-slotted strip waveguides exhibit weak overlap between the optical mode and the external electric field, resulting in the lowest $\Gamma/d$ values among the architectures discussed \cite{lu2020high}.\par 

Another metric of a ring modulator is assessed by the detuning efficiency $\eta_{MRM}$ given by \cite{eltes2020integrated, steglich2015novel,steglich2016partially}

\begin{equation}
	\eta_{MRM}=-\eta_{ps}\times\frac{\lambda }{\textit{n}_g \textit{L}_{ring}}\times\textit{L}_{ps}
\end{equation}

\noindent where $\uplambda$ is the operating wavelength, $\textit{L}_{ps}$ is the phase shifter length, $\textit{n}_g$ is the group index, and $\textit{L}_{ring}$ is the ring circumference. 
The most comprehensive FoM for the modulation efficiency of an MRM, however, can be obtained by \cite{lou2013design, xue2022breaking,takayesu2009hybrid}

\begin{equation}
	\eta_2 = \lambda_\text{off-res} / \eta_\text{MRM} \propto \textit{V}_{off-res}
\end{equation}

\noindent where $\lambda_\text{off-res}$ and $\textit{V}_{off-res}$ are the spectrum shift that switches the cavity to off-resonance condition to reduce the amplitude by maximum ER (e.g., 10 dB) and the voltage required for such a shift, respectively. Besides, $\eta_2$ depends on the Q-factor and the coupling condition of the ring. A higher Q results in lower $\textit{V}_{off-res}$. Usually, the phase shift required to achieve $\lambda_\text{off-res}$ is not sufficient; rather, a full 2$\pi$ shift is needed to shift the spectrum by an FSR and compensate for thermal fluctuations and fabrication imperfections. The voltage required to do so, i.e., $V_{\pi}$, is proportional to $V_{off-res}$, which is a more intuitive way of characterizing an MRM, whereas $\eta_\text{MRM}$ needs other information to describe how efficient the MRM is. Given the dependence of both $\textit{V}_{off-res}$ and the MRM's EO bandwidth on Q factor, design considerations are required to achieve the desired performance.

In fully slotted waveguide configurations, the resonator's quality factor, $\textit{Q}_{MRM}$, is notably affected by several loss mechanisms, including bend-induced radiation, sidewall roughness, absorption in the doped rails forming the slot, and the intrinsic loss of the OEO material \cite{christopoulos2019calculation}. Alternative designs have demonstrated the potential for achieving higher $\textit{Q}_{MRM}$ values, such as asymmetric slot structures \cite{anderson2006high} and partially slotted racetrack resonators \cite{steglich2016partially}.
Despite its demonstrated high modulation efficiency ($\eta_{MRM}$), this configuration remains limited by its EO bandwidth, the complexity of the control circuitry required due to the excessively high quality factor ($\textit{Q}_{MRM}$), and a marginally larger footprint. These factors collectively cast doubt on its suitability for applications requiring high integration density and broad bandwidth.\par

As an alternative approach, we implemented a semi-ridge, non-slotted waveguide within an MRM structure featuring a radius of $\textit{R} = 40~\mu m$, as depicted from various perspectives in Fig.~\ref{sims}. In this configuration, the optical mode exhibits asymmetric leakage beyond the undoped, non-slotted core waveguide (220 nm thick), which is coupled to a lightly doped slab layer ($\textit{t}_{slab}\approx$ 90 nm thick) on one side. This geometry facilitates the formation of an electric field between the core waveguide, the doped slab, and an overlaid metal electrode. The design aims to optimize the effective utilization of the electrodes, in contrast to conventional layouts that position electrodes laterally on both sides of a rib waveguide. Consequently, this configuration achieves an enhanced $\Gamma/\textit{d}$ ratio. The proposed structure requires one fewer doping step than a typical strip-loaded slot waveguide, which commonly involves three doping levels to balance electrical and optical losses \cite{ummethala2021hybrid}. Accordingly, the undoped core reduces doping-induced optical loss, a non-negligible limitation in slot waveguides \cite{ummethala2021hybrid}. Furthermore, the $5~\mu m$-wide oxide opening substantially mitigates infiltration of the organic electro-optic (OEO) material, compared to the narrow slot widths of $\approx 100 \sim 200$ nm typical of conventional slot waveguides.\par

Deep ultraviolet (DUV) lithography was employed at a commercial photonic foundry (Advanced Micro Foundry) to fabricate the devices on a standard silicon-on-insulator (SOI) wafer. The OEO material used was \textit{pm}-158, a proprietary FN-LC developed by Polaris Electro-Optics, with a representative molecular structure shown in Fig.~\ref{overview}(b) and (c). To ensure optimal interfacing with silicon, we implemented a chip surface-cleaning protocol that included an oxygen plasma treatment immediately before material deposition. This step was found to produce the most favourable surface conditions for FN-LC coating. The vial containing the FN-LC material was first heated to approximately $130^{\circ}$C for 3 minutes to restore the mixture's homogeneity, then cooled to around $70^{\circ}$C before a drop was extracted for deposition. This final step raises the droplet temperature above the ferroelectric nematic transition threshold. 
Simultaneously, the chip was heated to $\approx70^{\circ}$C for 60 seconds to facilitate proper infiltration while minimizing temperature gradients between the material and the chip surface. An FN-LC layer was formed to enable initial evaluation over a large area. As a precaution, a gentle nitrogen ($N_2$) flow was applied during coating to reduce oxidation; however, subsequent measurements indicated that this had a negligible effect. The material retained its viscosity and did not exhibit flow even at elevated temperatures, allowing for localized deposition at designated sites.\par 

\section{Device Functionalization and Characterization}

\begin{figure}[ht!]
\centering\includegraphics[width=1.02\textwidth]{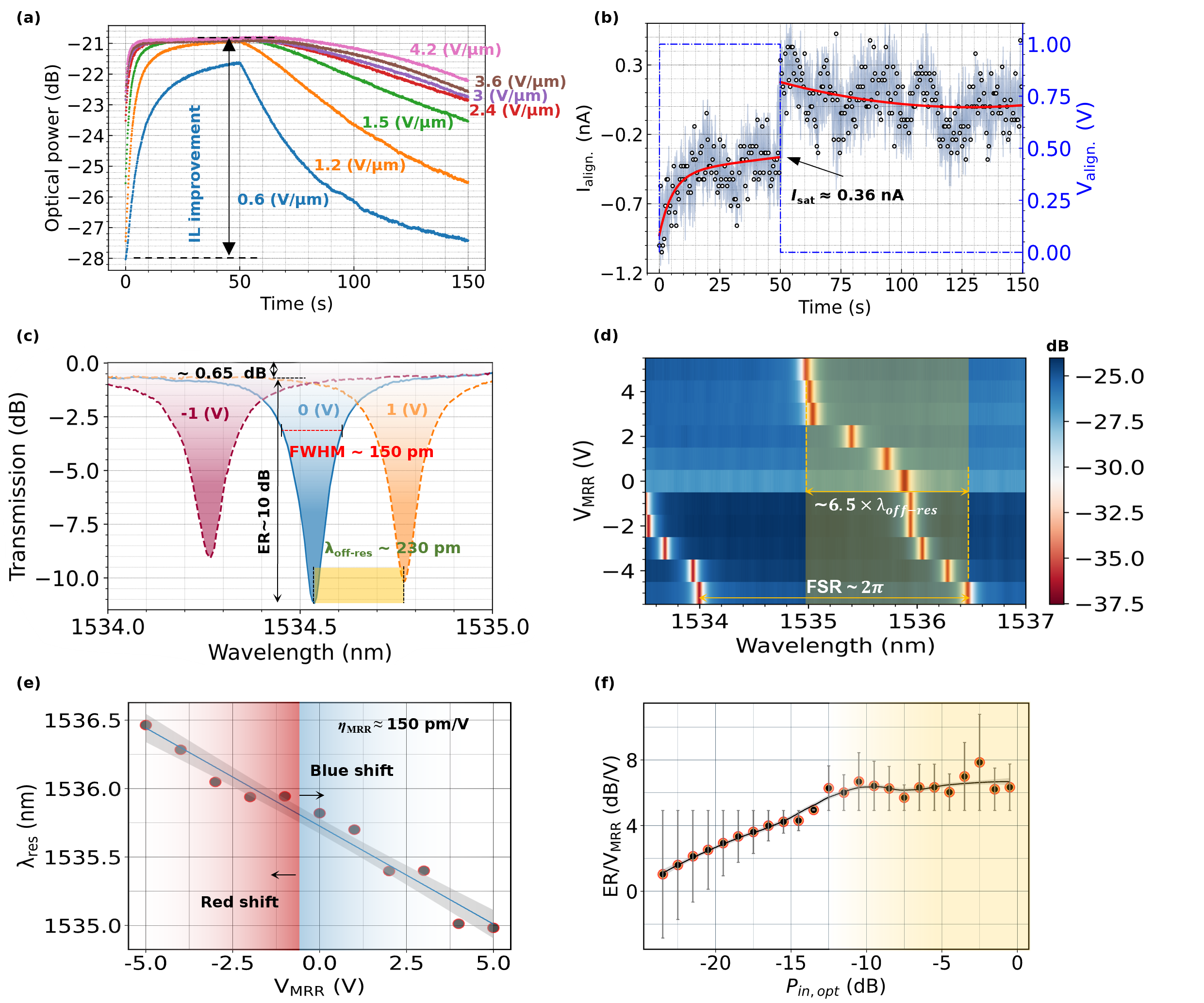}
\caption{FN-LC alignment and DC characterization. (a) Optical output power as a function of externally applied electric field (\textit{E}$_{ext}$). The insertion loss (IL) improves by $\approx$ 7 dB at a saturation field of $\textit{E}_{sat} \approx 1.2 V/\mu$m, corresponding to a DC voltage of $\textit{V}_{sat} \approx 2$ V, which is sufficient to achieve complete dipole alignment. (b) Polarization switching leakage current ($\textit{I}_{align}$) measured during the alignment process under \textit{V} = \textit{V}$_{sat}$. The current saturates for $\textit{t} \geq \textit{t}_{stat} = 50$ s, indicating that full alignment has been reached. A steady-state current of $\textit{I}_{sat} \approx 0.36$ nA is observed at $\textit{V}_{sat}$. Upon removal of $\textit{E}_{ext}$, a polarity reversal in $\textit{I}_{align}$ is detected, producing a modest back-switching current after $\textit{t}_{stat}$. (c) Normalized transmission spectrum at the through port for three different voltages ($\textit{V}_{MRM}$), demonstrating the ability to switch between on- and off-resonance states. A modulation efficiency of $\eta_{MRM} \approx 250$ pm/V (equivalent to approximately 1.67 FWHM/V), a quality factor of $\textit{Q}_{MRM} \approx 15340$, and an insertion loss of $\approx0.78$ dB are reported, where FWHM represents the full width at half maximum. (d) Two-dimensional heatmap of the transmission spectrum as a function of $\textit{V}_{MRM}$. Across the voltage range \textit{V} = -5 to 5 V, a cumulative resonance shift of approximately 1.5 nm ($\approx 6.5 \times \lambda_{off-res}$) is achieved. It corresponds to an estimated $\pi$-phase voltage of $V_{\pi} \approx 13.25$ V, required for a full $\pi$ phase shift equivalent to FSR/2, where FSR is the free spectral range. (e) Approximately-linear blue and red shift tunability measures a total wavelength detuning of $\approx 1.5\times$ FWHM/V over the range \textit{V} = -5 to 5 V. (f) Extinction ratio (ER) over $\textit{V}_{MRM}$ versus optical power injected to the MRM ($\textit{P}_{in,opt}$) shows minimal evidence of photo-oxidation at higher optical powers, confirming material stability under continuous optical excitation.}
\label{shift}
\end{figure}

The spontaneously polarized FN-LC phase requires an external electric field to orient the macroscopic dipole relative to the high-speed AC field used for modulation. This alignment step can be performed at room temperature, during regular RF operation, thus enabling a platform that functions without poling. However, to closely examine carrier dynamics and EO properties, we conducted this procedure separately, as depicted in Fig.~\ref{shift}(a-b). Compared to the results reported in \cite{taghavi2025ghz}, our observations demonstrate a more linear and smoother spectral transition as a function of the applied external electric field ($\textit{E}_\text{ext}$). The formation of a monodomain reduces insertion loss (IL) during dipolar alignment. Upon removal of the external field, the uniform alignment is gradually lost. 
Consequently, we anticipate that maintaining RF operation will require the application of a constant DC offset voltage.\par

Nevertheless, an external field as low as approximately $\textit{E}_{ext} \approx 1.2~V/\mu m$ was sufficient to achieve complete dipolar alignment, while the saturation current remained minimal at approximately $\textit{I}_{sat} \approx 0.36~nA$. When aligning multiple devices at scale, this negligible LCT limits local heat generation, which can be further mitigated by incorporating a CBL. 
During measurements, we observed that the alignment step does not require an encapsulation coat or an inert gas atmosphere, as no oxidation-induced deterioration was noted \cite{xu2020ultrahigh,anderson2014mechanisms}. This durability is partly due to the step being conducted at room temperature, which simplifies the in-situ measurement of the optical output and, hence, the direct assessment of $\textit{r}_{33}$ \cite{olbricht2008laser}. Nevertheless, the material has been designed to be oxidatively more stable.\par

Fig.~\ref{shift}(c) proves that the MRM has a substantial phase shift mechanism with only a $\textit{V}_{pp}\approx1~V$ value predominantly governed by the material's birefringence effect. 
We studied the device's spectrum at $V\gg V_{pp}$ to create the spectrum shift required to fully move from the resonance condition (i.e., $\lambda_{off-res}\approx0.23~nm$) as shown in Fig. \ref{shift}(c) aimed to investigate the necessary voltage to shift a full free-spectral range (FSR), i.e., $2\pi$. The device demonstrated the ability to maintain an approximately linear transfer function with an efficiency of $\eta_\text{MRM} \approx 150$ pm/V, sustained over a tuning range equivalent to $\approx 6.6\times \lambda_{off-res}$ or $\approx 10 \times \text{FWHM}$ as illustrated in Fig.~\ref{shift}(e), where FWHM represents the full width at half maximum. This characteristic is particularly compelling for applications such as neuromorphic photonic circuits \cite{singh2022neuromorphic}. In contrast to other specific modulation mechanisms, the observed blue and red spectral shifts in response to negative and positive voltages, respectively, suggest improved power consumption.

Further analysis of the device transfer function across a broader $V_\text{MRM}$ range yielded a half-wave voltage of $V_{\pi} \approx 13.25$ V, indicating a reduction in $\eta_\text{MRM}$ due to saturation effects within the transfer function.
This metric corresponds to a voltage-length product of $\textit{V}_{\pi}\textit{L}_{ring} \approx 3.33~\text{V} \cdot \text{mm}$, leading to a calculated static power consumption of $\textit{P}_{stat} \approx \textit{V}_{\pi} \times \textit{I}_{\text{stat}} \approx 4.77~\text{nW}/\uppi$.
Despite the acquired waveguide's non-slotted nature, this high modulation efficiency is attributed to the strong birefringence and the dielectric properties of FN-LC. The observed fluctuations in Fig.~\ref{shift}(e) are also likely due to the mentioned high sensitivity of the resonance wavelength to the employed 1~V step voltages. Nevertheless, one might argue that the issue should not be a concern in practice, as the effect should apply only to DC changes. Reducing the step voltage to lower values partially mitigated the nonlinearities in response to this slowly varying bias.\par

We also conducted a preliminary aging test to measure the material decay in the presence of optical power. As seen in Fig.~\ref{shift}(f), no substantial decrease in ER/\textit{V}$_{MRM}$ as a function of laser power was observed during the course of a few hours of this measurement. Even though more rigorous aging tests are required to address concerns about photobleaching or photooxidation, the proposed FN-LC MRM still offers robust performance compared to other OEOs we investigated.  
One possibility is that $10\sim50~\mu m$ material on top protects the $\sim$500 nm underneath where the optical mode exists.\par

\begin{figure}[ht!]
\centering\includegraphics[width=\textwidth]{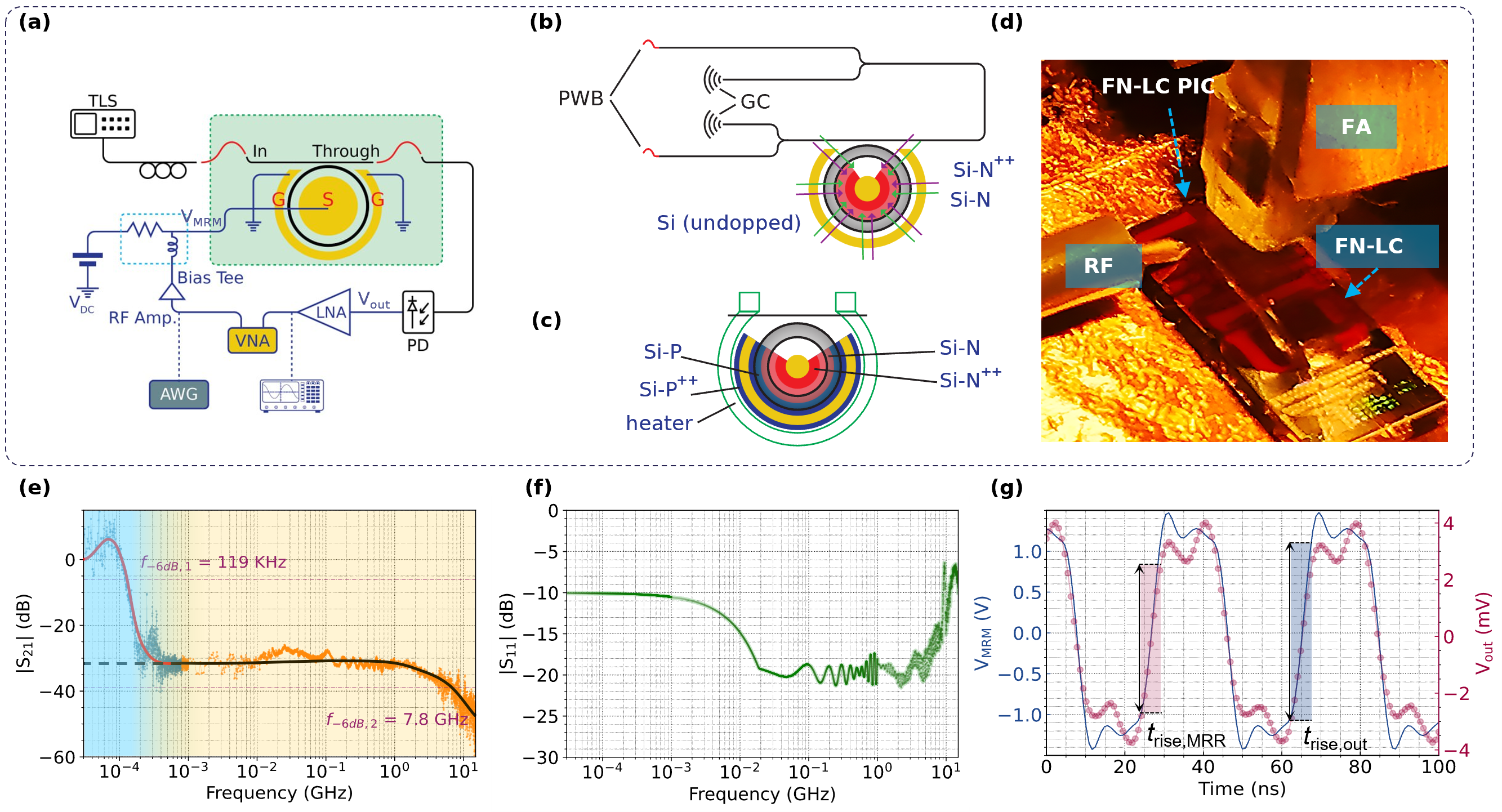}
\caption{AC characterization. (a) Schematic of the experimental setup for high-frequency characterization. A vector network analyzer (VNA) is employed for \textit{S}-parameter measurements. Each microring modulator (MRM) is fed an optical carrier via either grating couplers (GCs) or photonic wire bonds. For time-domain analysis, the VNA is replaced by an arbitrary waveform generator (AWG) and an oscilloscope. (b) The proposed device architecture requires only a single dopant type and two implantation levels, in contrast to (c) a conventional \textit{pn}-junction-based MRM, which utilizes three implantation doses for both \textit{P}- and \textit{N}-type dopants. The birefringence-induced phase shift in ferroelectric nematic liquid crystal (FN-LC) replaces the power-intensive thermo-optic heater, thereby enhancing the extinction ratio. The RF field (purple arrows) is applied parallel to the pre-set molecular alignment (green arrows). (d) A photonic integrated circuit comprising active devices coated with 10$\sim$50 nm thick FN-LC during device-level testing. (e) Measured $\textit{S}_{21}$ and (f) $\textit{S}_{11}$ parameters of the MRM under test. Two distinct phase-shift mechanisms are identified: a slow yet strong response and a faster yet weaker one, with corresponding cutoff frequencies ($\textit{f}_{-6dB}$) of 119 kHz and 7.8 GHz, respectively. These are differentiated from the previously observed substantial DC phase shift. The amplitude difference between the two response trends ($|\Delta S_{21}|\approx30~\mathrm{dB}$) suggests two separate phase modulation processes characterized by differing $\textit{r}_{33}$ values across frequency bands. (g) Time-domain response of the MRM to a 25 MHz square wave, exhibiting minimal distortion and thereby validating the presence of the Pockels effect in the FN-LC material.}
\label{AC}
\end{figure}

The key advantage of the FN-LC lies in its ability to respond to rapidly varying electrical signals—an area in which otherwise comparable PN-LC-based devices typically fall short. To evaluate this capability, the device was tested using the configuration depicted in Fig.~\ref{AC}(a). A microwave analyzer (N9917A FieldFox, Keysight), followed by a microwave amplifier (SHF-s807), was employed to drive the MRM with $\textit{V}_{pp} \approx 2.83$ V over a frequency range spanning 30 kHz to 11 GHz. Device biasing was provided via a precision source measure unit (2600B Series, Keithley).\par

The observed $\eta_\text{MRM}$ at DC was sufficiently high to eliminate the need for an additional power-intensive thermal heater, which is typically required in \textit{pn}-junction-based ring modulators, as shown in Fig.~\ref{AC}(c). Beyond reducing the number of electrical I/Os required to drive the resonator, the exceptionally low static power consumption, $\textit{P}_{stat}$, represents a substantial advantage of the FN-LC material platform—particularly for applications involving large-scale PICs populated with numerous ring resonators, such as switch banks.\par  

Two types of I/O were used: an optical 4-channel fibre array (8-degree polished angled FA by OZ Optics) and an RF probe (Picoprobe Model 40A by GGB). The optical carrier was generated by a tunable laser source (8164A/81682A, Agilent). It was picked up by an ultrafast fibre-optic photoreceiver (RXM40AF by Thorlabs) and amplified by a low-noise amplifier (PSPL5828A by Picosecond). The frequency and time domain responses shown in Fig.~\ref{AC}(d-f) indicate the presence of at least two phase shift mechanisms with an EO bandwidth of $\textit{f}_\text{-6dB}\approx~119~KHz$ and $\approx~7.8~GHz$. The former is more consistent with commonly reported switching times for PN-LC \cite{xing2015digitally}, which are attributed to dipolar orientation. At the same time, the latter confirms the Pockels phase shift mechanism observed in FN-LC \cite{taghavi2025ghz}. The considerable difference in magnitude and cut-off frequency between the two shown roll-offs confirms the existence of two distinct phase-shift processes. In \cite{onural2024hybrid}, researchers reported EO bandwidth up to 35 GHz for a slot waveguide MZM, proving that the material's intrinsic bandwidth is not the limiting factor. 
Instead, the bandwidth is constrained by the proposed device's structure (i.e., resonant-based) and its relatively high RC time constant for the equivalent circuit. Nevertheless, given the highest EO bandwidth of a SOH MRM reported for Pockels-based organic materials so far is $\approx1.34~GHz$ \cite{gould2011silicon}, our proposed device reports a higher record, primarily thanks to the non-slotted waveguide used in this work.\par
	

The device was subjected to elevated optical laser ($\approx$ 8 dBm) and electrical RF ($\textit{P}_{in,elec} = 13$ dBm) power levels, exceeding those used during standard operation. The ambient temperature was varied from 5$^{\circ}$C to 55$^{\circ}$C in 5$^{\circ}$C increments, with each temperature maintained for 3 days. Subsequent microscopic inspection revealed no visible signs of degradation, and $|\textit{S}_{21}|$ remained unchanged, demonstrating the resilience of FN-LC-induced phase shifts against common aging factors such as oxygen and humidity. To evaluate the device's suitability for switching applications, time-domain measurements were conducted. These revealed minimal signal distortion and no observable variation in rise or fall times, as illustrated in Fig.~\ref{AC}(g). Although the extinction ratio (ER) at 7.8 GHz was relatively modest, the device is expected to achieve exceptionally high dynamic energy efficiency, with an estimated $\textit{E}_{\text{dyn}} \approx 1.65$ \textit{f}J/bit, owing to its intrinsically low capacitance.\par 

\begin{figure}[ht!]
\centering\includegraphics[width=\textwidth]{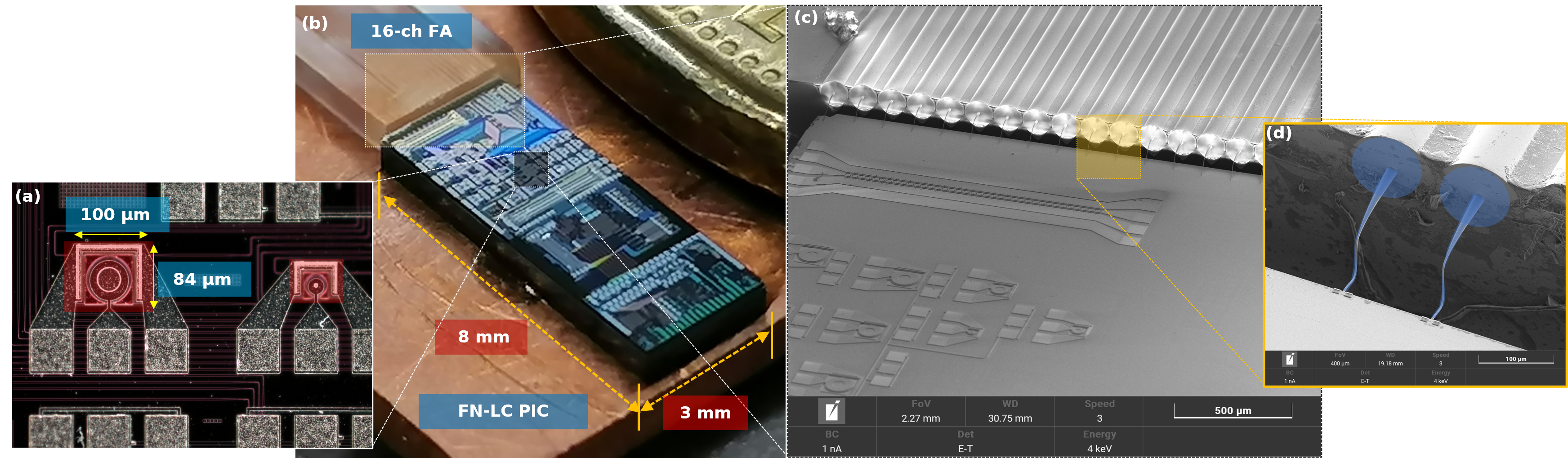}
\caption{Implemented photonic integrated circuit (PIC) prototype incorporating ferroelectric nematic liquid crystal (FN-LC). (a) Microscopic top-down image of the fabricated microring modulator coated with FN-LC, exhibiting an overall footprint of $\approx8400~\mu m^2$. The image highlights the individual electrical pads and RF routing paths used for both device-level and chip-level characterization. (b) Close-up view of the PIC, showcasing a variety of integrated components within a compact 24 $mm^2$ area. A 16-channel, zero-angle fiber array (FA) delivers the optical carrier via photonic wire bonds (PWBs). FN-LC is selectively deposited on active regions, thereby mitigating unintentional changes in the refractive index in other areas exposed through the oxide open window. (c) Scanning electron microscopy (SEM) image of the PIC prior to FN-LC deposition, with an intentional metallic coating applied to prevent charge accumulation during imaging. (d) Magnified SEM image of two representative PWBs, illustrating bend radii. The fiber tips in the FA V-groove are aligned and bonded to the corresponding on-chip edge couplers via PWBs.}
\label{packaging}
\end{figure}

To evaluate the suitability of the proposed FN-LC-based silicon-organic hybrid (SOH) platform for large-scale applications, a PIC comprising 58 modulators of various types, including the MRM presented in this study, was fabricated. To avoid the limitations of chip-scale optical I/O using other conventional techniques, either via edge coupling or glued fiber array (FAs), PWBs were established between the fiber facets of a 16-channel FA with a 0° polished angle and the on-chip edge couplers to facilitate optical input/output at the chip level. Precise manual alignment between the on-chip alignment markers and the core of each FA channel ensures robust PWBs with coupling efficiency. Optimization of optical coupling losses between the FA and the PIC was performed by minimizing both vertical and horizontal offsets at the bond interfaces, thereby reducing mode mismatch and bending losses. The average coupling loss is $\approx$12 dB per channel (compared to $\approx$ 10.5 dB per grating couplers (GC)), with the potential for reduction to $\approx$2 dB as reported in \cite{lin2023cryogenic}. Potential steps to improve coupling loss included better height match and further bond optimization between the fiber and edge couplers. The PWB process used two-photon polymerization (Sonata 1000, Vanguard Automation \cite{lindenmann2012photonic}) to fabricate three-dimensional, freeform optical waveguides. These structures facilitate efficient coupling of light from the FA into the PIC. Representative images of the FN-LC-integrated PIC are provided in Fig.~\ref{packaging}.

A subset of the test devices was manually coated using a capillary glass tube, which was positioned under a probe station using a micropositioner. One potential challenge involves determining the optimal sequence of PWB and FN-LC coating. In the demonstrated PIC, selective coating of devices was performed as the final step in the fabrication process, since FN-LC is soluble in the solvent used during PWB formation. A limitation of this approach is that the surface passivation step had to be omitted to preserve the integrity of the PWB. This omission may have adversely affected the infiltration of material into exposed oxide regions, thereby limiting LMI efficiency.

\section{Discussion, optimization and future outlook}

This work presents the advantages of a CMOS-compatible silicon-organic hybrid (SOH) platform based on FN-LC. By confirming the coexistence of two distinct phase-shifting mechanisms \textemdash~a GHz-speed, Pockels-effect-based response and a slower, yet highly efficient, birefringence-driven response \textemdash we report the first demonstration of an MRM selectively coated with FN-LC. The device features a 5 $\mu$m-wide oxide opening that allows proper infiltration of FN-LC, unlike relatively challenging sub-200 nm slotted waveguides, which often fail to fill with the EO material. Our proposed device utilizes an enhanced overlap between optical and electrical modal fields brought by electrode placement. Additionally, it avoids excessive waveguide doping, which does not limit the device's EO bandwidth but improves detuning efficiency by increasing the quality factor.\par

At DC, we achieved a detuning efficiency of at least 150 pm/V, corresponding to a static power consumption of approximately $4.77~\textit{nW}/\uppi$ and a modulation efficiency of $V_{\pi}L \approx 3.33~V\cdot mm$. This intrinsic, energy-efficient tuning mechanism outperforms conventional \textit{pn}-junction-based technologies, which typically rely on power-intensive thermo-optic heaters to achieve resonance tuning \cite{sun2019128}. One key advantage of this efficient tuning method is the ability to merge the RF and DC drive lines, which is helpful for integration at larger scales. Of particular interest is the non-thermal nature and low power consumption of this slow yet efficient DC detuning, which eliminates the need for thermal cross-talk for densely integrated switch networks and MRR weight banks \textemdash~ a prevalent issue in conventional technology based on \textit{p-i-n} devices.

\begin{figure}[ht!]
\centering\includegraphics[width=0.8\textwidth]{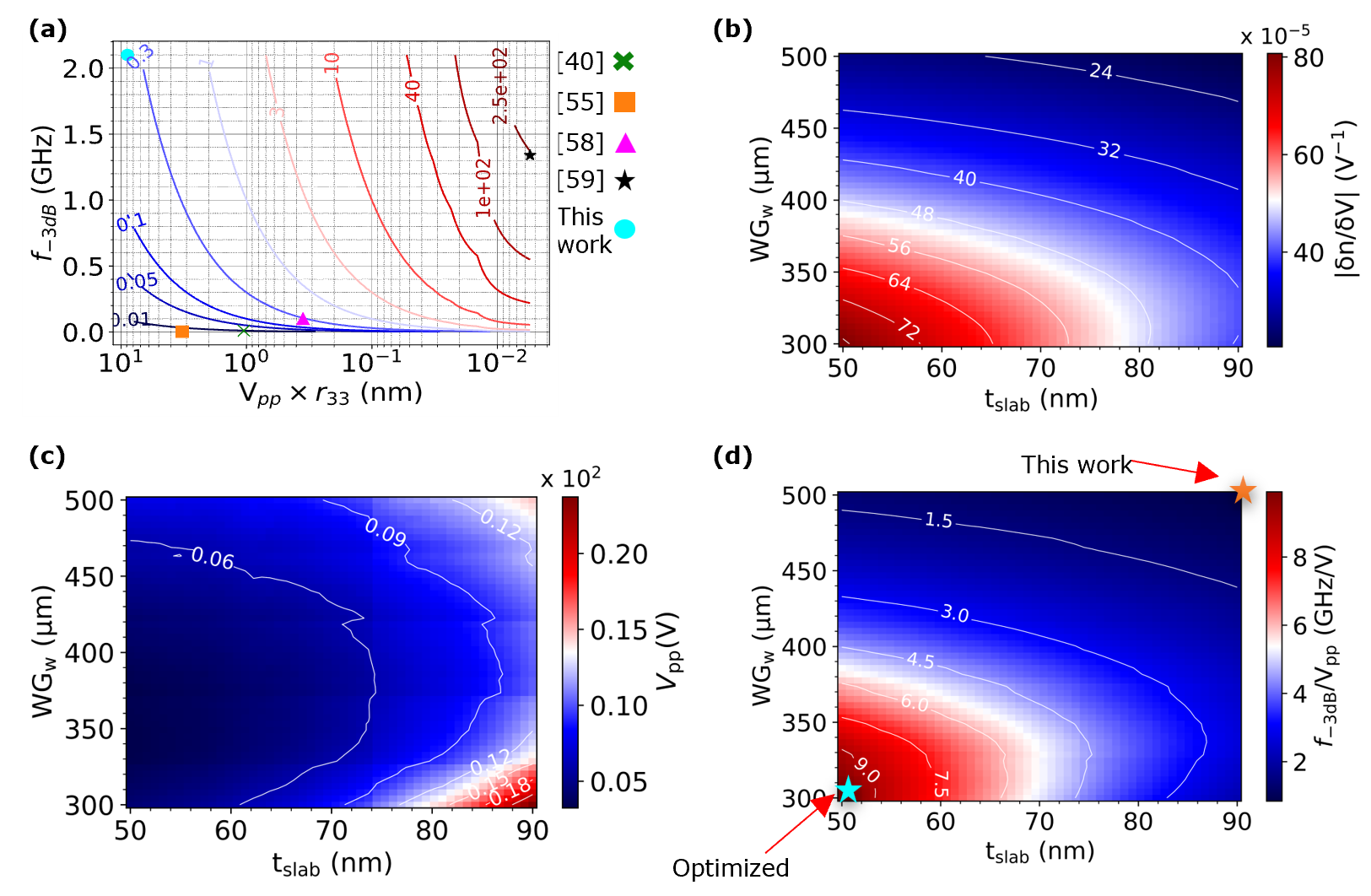}
\caption{Comparison and optimization. (a) Benchmarking of state-of-the-art silicon organic hybrid (SOH) microring modulators against the present work, based on two key figures of merit: modulation bandwidth and drive voltage, both normalized with respect to the Pockels coefficient ($\textit{r}_{33}$) of the employed organic material. (b-d) Simulated optimization of waveguide susceptibility ($\eta_{ps}$), peak-to-peak drive voltage ($\textit{V}_{pp}$), and the ratio $\textit{f}_{-3dB}/\textit{V}_{pp}$ as functions of device structure (i.e., $\textit{WG}_w$ and $\textit{t}_{slab}$). Compared to the experimentally obtained values from the current device, $\eta_{ps} \approx 0.69 \times 10^{-5}$ V$^{-1}$, $\textit{V}_{pp} \approx 368$ V, and $\textit{f}_{-3dB} \approx 2.1$ GHz, these structural optimizations are predicted to yield substantial performance enhancements, potentially achieving $\textit{V}_{pp} < 5$ V and $\textit{f}_{-3dB} > 40$ GHz due to a significantly improved waveguide susceptibility ($\eta_{ps} \geq 80 \times 10^{-5}$ V$^{-1}$).}
\label{optim}
\end{figure}

At AC, a high-speed EO response was confirmed through \textit{S}-parameter analysis, revealing a device-limited EO bandwidth of at least $\textit{f}_{-6dB} \approx 7.8~\text{GHz}$. This result establishes a new performance benchmark for SOH ring modulators, as summarized in Fig.~\ref{optim}(a). Despite a conservative electrode-to-waveguide spacing, the device operates at a relatively low driving voltage ($\approx2.83 $ V), obviating the need for a dedicated high-voltage driver.\par

From a large-scale integration perspective, the FN-LC platform eliminates the need for electro-thermal poling, a step required for EOP-based devices, thereby simplifying the control of individual devices. Due to the substantial spontaneous polarization of FN-LC, established at room temperature, the required electric field strength is approximately 50 to 100 times lower than that of typical EOPs (e.g., \cite{xu2022design}) when applied to an equivalent device structure. Additionally, the high viscosity of FN-LC facilitates selective coating and enhances alignment precision during fabrication.\par 

In this work, we prioritized low drive voltage, CMOS compatibility, and fabrication simplicity over maximizing EO bandwidth. The Pockels coefficient of the FN-LC used in this work is lower than that of commercially available EOPs \cite{taghavi2025ghz}, which designates a promising candidate for future enhancement through material engineering. Additionally, the EO bandwidth of the present device is lower than that of state-of-the-art plasma-dispersion-based silicon modulators. This limitation is not a blocker and should not overshadow the broader advantages of this emerging hybrid material platform. Improvements in device design could address the relatively modest EO bandwidth while preserving other performance metrics, albeit at the expense of increased fabrication complexity. \par

\begin{figure}[ht!]
\centering\includegraphics[width=\textwidth]{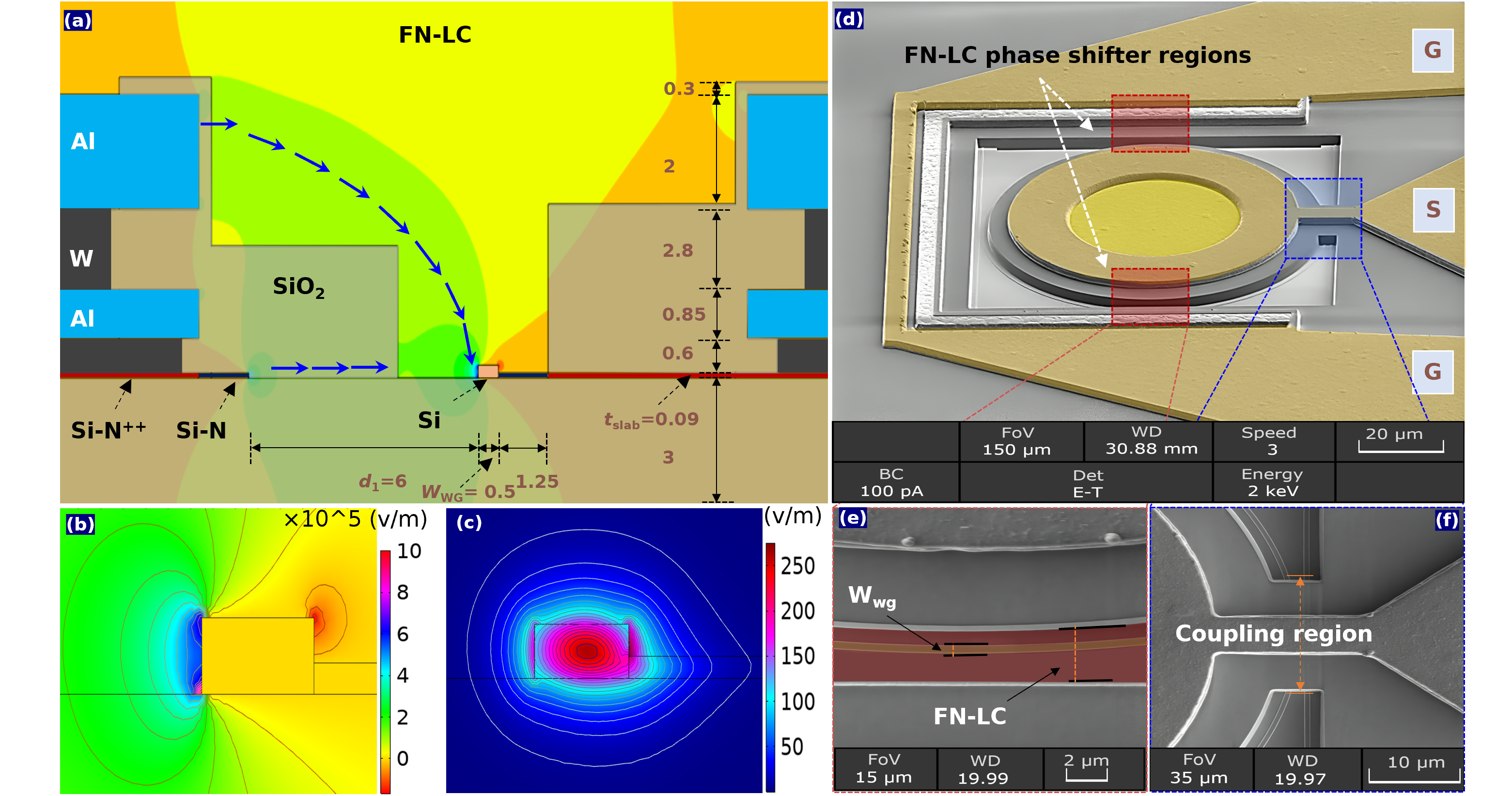}
\caption{Device structure and design. (a) The layer stack-up of the microring modulator (MRM) (all dimensions are in $\mu m$) overlapped with the electric field ($\textbf{E}_e$). The partial etch layer (Si-N with \textit{n}-type dopant implanted at a dose of 3$\times10^{17}~cm^{-3}$) is attached to the full-etch waveguide (Si undoped) on one side and to the highly doped interface (Si-N$^{++}$ with \textit{n}-type dopant implanted at a dose of 1$\times10^{20}~cm^{-3}$) for ohmic contact on the other side. The single-sided rib waveguide reduced the effective distance between the anode and cathode, increasing the device factor ($\Gamma/\textit{d}$) and detuning efficiency compared to an equivalent non-slotted waveguide. $\textbf{E}_e$ and transverse electric optical field ($\textbf{E}_x$) overlap each other from the left and top-left. (b) $\textbf{E}_e$ and (c) normalized $\textbf{E}_x$ in y-slice view. (d) A false-colored scanning electron microscopy (SEM) image of the constructed MRM demonstrating the two-phase shifter regions with the shortest distance between the inner (signal or "S") and outer (ground or "G") electrodes, where $\Gamma/\textit{d}$ is maximized. SEM image of (e) one of the phase shifter regions shows the oxide open area for ferroelectric nematic liquid crystal infiltration, and (f) the coupling region between the waveguide bus and microring running under a metallic bridge carrying the RF signal.}
\label{sims}
\end{figure}

Several design modifications could further reduce $\textit{V}_{pp}$ at both DC and AC. For example, replacing the rectangular outer electrode geometry with a horseshoe-shaped design would increase the effective interaction length along the ring circumference. Additionally, reducing the electrode-to-core distance ($\textit{d}_1$) and removing the upper oxide cladding layer on the slab region (see Fig.~\ref{sims}(a)) would enhance the EO overlap factor, $\Gamma$.\par

Modifying the semi-ridge waveguide geometry, specifically, the waveguide core width ($\textit{W}_{wg}$) and the slab thickness ($\textit{t}_{slab}$) can also improve $\eta_{ps}$, while preserving the \textit{Q}-factor. Our simulations indicate that removing the left oxide cladding and introducing even a light doping (e.g., $1\times 10^{15}$ cm$^{-3}$) to the waveguide core can reduce the effective electrical resistance (\textit{R}) from $\approx 1.2~\text{M}\Omega$ to $\leq40~\text{k}\Omega$.
~The 3-dB bandwidth ($\textit{f}_{-3dB}$) would then be primarily limited by the loaded quality factor, according to:

\begin{equation}
	\textit{f}_{-3dB}^{-2}=\big[2\pi R C\big]^2+\big[\frac{\lambda Q_{MRM}}{c}\big]^2
\end{equation}

\noindent Here, \textit{c} denotes the speed of light, and \textit{C} ($\approx0.63$ \textit{f}F) represents the equivalent capacitance of the MRM. The proposed design optimizations position the MRM reported in this work on a more competitive trajectory relative to alternative technologies. A summary of potential enhancements applicable to the current device architecture is presented in Fig.~\ref{optim}(b-d), where a next-generation MRM design is projected to achieve a 3-dB bandwidth exceeding 40 GHz and a peak-to-peak driving voltage below 5 V.\par 

Despite the colossal permittivity of FN-LC at DC ($\upvarepsilon_{FN-LC}$), introducing an atomically-thin CBL made of high-k material (e.g., $TiO_2$) at the FN-LC/silicon waveguide and FN-LC/aluminum electrode interfaces may help preserve dipole alignment by inhibiting LCT, which is mainly contributed by the polarization reversal current through the FN-LC layer between the two electrodes. Indeed, the voltage drop across the CBL at DC can be estimated by

\begin{equation}
		\frac{\textit{V}_{FN-LC}}{\textit{V}_{CBL}}\approx2\frac{\upvarepsilon_{CBL}}{\upvarepsilon_{FN-LC}}\times \frac{\textit{d}_{FN-LC}}{\textit{d}_{CBL}}
\end{equation}

\noindent where $\upvarepsilon_{CBL}$) is the permittivity of the CBL, $\textit{d}_{FN-LC}$ (i.e., $\approx \textit{d}_1$ in Fig.\ref{sims}(a)) and $\textit{d}_{CBL}$ is the thickness of the FN-LC and CBL, respectively. While at DC $\upvarepsilon_{FN-LC}$ surpasses $10^4$ \cite{li2021development}, but since $\textit{d}_{FN-LC} \ll \textit{d}_{CBL}$, the majority of $\textit{E}_{ext}$ still drops across the highly conductive but thick FN-LC layer. The speed degradation due to the parasitic capacitance introduced by adding such a CBL is also insignificant, as the permittivity of FN-LC sharply drops at frequencies above MHz.

Besides, in EOP counterparts, degradation due to environmental exposure to oxygen and humidity often necessitates the use of encapsulation strategies \cite{lee1995evidence}. Although no signs of material degradation were observed during our measurements, similar protective measures could be considered for FN-LC devices. To obtain a more comprehensive understanding of the material's aging characteristics, both standard and accelerated aging tests-such as exposure to 85 $^{\circ}$C under nitrogen, ambient air, or 85$\%$ relative humidity (RH) in damp heat conditions \cite{hammond2022organic}-are essential.

For scalable and more efficient device fabrication, inkjet printing offers a promising alternative to the manual coating method used in this study. This technique enables selective deposition of FN-LC over high-density active components, leveraging the material's inherent high viscosity. In addition to potential anti-aging advantages, applying masking layers (e.g., using polydimethylsiloxane) to coated devices can enhance their resilience during the PWB process. This approach is also favourable for minimizing PWB insertion loss and improving surface preparation before FN-LC infiltration.
The next logical step is to develop a fully integrated PIC with spatially localized FN-LC coatings. Such an architecture would be particularly suitable for high-density, reconfigurable photonic systems (e.g., for weight banks), optical transducers, and switching networks.

\par
	
	


\begin{backmatter}
\bmsection{Funding}
This research was funded by the Natural Sciences and Engineering Research Council of Canada (NSERC), the SiEPICfab consortium, the B.C. Knowledge Development Fund (BCKDF), the Canada Foundation for Innovation (CFI), MITACS, and Schmidt Sciences.

\bmsection{Acknowledgment}
The authors acknowledge Omid Esmaeili's helpful insights into device characterization. They are also grateful to Matthias Kroug for his guidance on SEM images and Hang Bobby Zou for his help with PWB assembly.

\bmsection{Disclosures}
The authors declare no conflicts of interest.

\bmsection{Data Availability Statement}
Data underlying the results presented in this paper are not publicly available at this time but may be obtained from the authors upon reasonable request.

\end{backmatter}

\bibliography{sample}

\newpage

\end{document}